\documentclass[letterpaper]{article}
\pdfoutput=1

    \PassOptionsToPackage{numbers, compress}{natbib}
\usepackage[final]{neurips_2023}

\usepackage[utf8]{inputenc} % allow utf-8 input
\usepackage[T1]{fontenc}    % use 8-bit T1 fonts
\usepackage{booktabs}       % professional-quality tables
\usepackage{amsfonts}       % blackboard math symbols
\usepackage{nicefrac}       % compact symbols for 1/2, etc.
\usepackage{microtype}      % microtypography
\usepackage{xcolor}         % colors
\usepackage{graphicx}
\usepackage{enumitem}
\usepackage{mdframed}
\usepackage{float}
\usepackage{graphicx}
\usepackage[affil-it]{authblk}
\usepackage[numbers]{natbib}
\usepackage{hyperref}       % hyperlinks
\usepackage{url}            % simple URL typesetting

\title{Towards Publicly Accountable Frontier LLMs\\[0.7ex]\small Building an External Scrutiny Ecosystem under the ASPIRE Framework}

\author[1,2,*]{Markus Anderljung}
\author[1]{Everett Thornton Smith}
\author[1,3]{Joe O’Brien}
\author[1]{Lisa Soder}
\author[1]{Benjamin Bucknall}
\author[1]{Emma Bluemke}
\author[1]{Jonas Schuett}
\author[1,4]{Robert Trager}
\author[5]{Lacey Strahm}
\author[6,7]{Rumman Chowdhury}

\affil[1]{Centre for the Governance of AI}
\affil[2]{Center for a New American Security}
\affil[3]{Institute for AI Policy and Strategy }
\affil[4]{Blavatnik School of Government, University of Oxford}
\affil[5]{OpenMined}
\affil[6]{Humane Intelligence}
\affil[7]{Harvard Berkman Klein Center}
\affil[*]{Corresponding Author:  markus.anderljung@governance.ai}

\begin{document}

\maketitle

\begin{abstract}
With the increasing integration of frontier large language models (LLMs) into society and the economy, decisions related to their training, deployment, and use have far-reaching implications. These decisions should not be left solely in the hands of frontier LLM developers. LLM users, civil society and policymakers need trustworthy sources of information to steer such decisions for the better. Involving outside actors in the evaluation of these systems – what we term "external scrutiny" – via red-teaming, auditing, and external researcher access, offers a solution. Though there are encouraging signs of increasing external scrutiny of frontier LLMs, its success is not assured. In this paper, we survey six requirements for effective external scrutiny of frontier AI systems and organize them under the ASPIRE framework: Access, Searching attitude, Proportionality to the risks, Independence, Resources, and Expertise. We then illustrate how external scrutiny might function throughout the AI lifecycle and offer recommendations to policymakers.
\end{abstract}

\section{Risks from Frontier LLMs}
The most capable large language models (frontier LLMs) \cite{govuk} pose significant risks of harm, both now and in the future \cite{Weidinger_2022}. These risks include LLMs exacerbating discrimination \cite{id1}, disinformation and election interference \cite{id2}, authoritarian or corporate surveillance \cite{id1}, cyberattacks \cite{id4}, or the proliferation of weapons of mass destruction, especially biological weapons \cite{id5}. However, these risks are not inevitable; they are a function of decisions around model development, deployment and use.

\section{AI Governance as an Information Problem}
Just as risk of harm warrants governance in industries as diverse as aviation, pharmaceuticals, finance, and nuclear power, so too do the risks from articifial intelligence (AI). However, in order to make informed decisions and design well-targeted policies, stakeholders outside of AI developers need reliable information about frontier LLMs \cite{Whittlestone, Brundage}.  As such, good governance is in part an information problem \cite{id6-1, id6-2, id6-3}. When users, civil society and policymakers possess reliable information, they can contribute to the responsible development, deployment, and use of frontier LLMs in a variety of ways: 
	
\begin{itemize}
    \item \textbf{Users. }With a firmer understanding of LLMs, users can make better choices about what systems to use and pressure AI developers to act more responsibly \cite{id7-1, id7-2, id7-3}.
    \item \textbf{Civil society. }More accurate information about LLMs can allow civil society to better research and advocate for policies, standards, or other methods to reduce harm \cite{id7-1, cihon}.
    \item \textbf{Policymakers. }By better understanding the potential risks and impacts of AI – present and future – policymakers can legislate and regulate more effectively.
\end{itemize}

While reliable information alone is far from sufficient for good governance, it is close to a necessary condition. It is hard to imagine how a society could ensure that its pharmaceuticals, airplanes, financial system, or nuclear power plants were safe and dependable without a reliable source of information about them. The same is true for AI.

\section{Sourcing Reliable Information}
AI developers are the primary party responsible for ensuring their systems are safe \cite{id22}, and the risk assessments they perform and other information they provide are a valuable resource for policymakers \cite{id54}. However, policy decisions must not solely rest on information provided by AI developers.

External actors – actors not employed by an AI developer – must be involved in the generation and distribution of risk assessments and other information necessary for good governance. This involvement of external actors is a process we call “external scrutiny.” Such scrutiny can involve auditing \cite{id8, id7-1, goodman}, evaluations \cite{id19-1}, red-teaming \cite{Brundage}, or other research performed on an AI system by e.g. government agencies, academics, nonprofits, or private companies. By doing so, control over the impacts of frontier LLMs can be "democratized" \cite{id7-3}.

We believe three types of information about frontier LLMs are particularly important for good governance \cite{id8}:
\begin{itemize}
    \item \textbf{Model capabilities.} What is the model capable of? What tasks can it perform? This can inform appropriate use-cases, as well as identify the model’s potential for misuse \cite{id19-1}.
    \item \textbf{Model controllability.} Does the model reliably act in accordance with user or developer intentions (e.g. does it produce unintentional toxic content \cite{tox}, or behave predictably out of distribution)? This informs the likelihood that the model will cause unintended harm.
    \item \textbf{Model impacts.} What effects might the model have on the world (e.g. exacerbating disinformation, or displacing workers) \cite{id1}? This can inform regulator or AI company decisions to introduce additional safeguards or even to recall a deployed model.
\end{itemize}

External scrutiny can help provide more reliable information by:
\begin{itemize}
    \item \textbf{Verifying developer claims. }Developers might misrepresent what they know about their LLMs. In the tobacco industry, companies had data that their products were harmful, but concealed evidence of harm and made fraudulent claims about product safety for approximately fifty years \cite{id9}. AI developers will face similar incentive problems when informing stakeholders about risks from their products. External scrutiny can reduce such information asymmetries between AI developers and external actors \cite{Brundage, cihon}.
    \item \textbf{Uncovering new information.} Developers may fail to identify issues with their systems. They may have incentives to put blinders on and remain unaware of significant issues. This problem is exacerbated by the vast space of potential model behaviors and uses, and the fact that understanding the inner workings of LLMs is notoriously difficult; even if they earnestly tried, AI developers are unlikely to uncover all important system flaws. External scrutiny can overcome these challenges by bringing a wider set of expertise, perspectives, and motivations to bear on the problem of identifying risks and issues with LLMs.
\end{itemize}

\section{Calls for External Scrutiny}
In recognition of these benefits, there are many efforts to expand external scrutiny of frontier LLMs. The United Kingdom has established the Frontier AI Task Force with £100 million in funding to ensure that AI risk assessments are conducted by neutral third parties \cite{id11, ft}.  In the United States, 15 leading AI companies have signed onto the White House Voluntary AI Commitments, which call for external red-teaming of AI models \cite{id12}. There are also several related proposals in the US Congress \cite{id14-1, id14-2, id14-3}. In the European Union, the latest parliamentary proposal on the AI Act would require involvement of “independent experts” in the design and testing of foundation models \cite{id16}. AI developers have also begun voluntarily inviting external parties to evaluate their systems \cite{id43,id17-1,id17-2, id17-4}.

Though developing quickly, the external scrutiny ecosystem for frontier LLMs is nascent and its effectiveness will depend on specific decisions. In the next section, we outline what some of these decisions are and offer a guide for navigating them.

\section{Requirements for Effective External Scrutiny}
Successfully designing and implementing external scrutiny of frontier LLMs will be a difficult task, with many potential pitfalls. If designed or implemented poorly, external scrutiny might not only fail to provide crucial information to reduce risks from AI, but could also create a false sense of security.

The collapse of Enron provides a notorious example of failed external scrutiny. Enron’s external auditors failed to uncover and report ongoing financial fraud due to a lack of independence from company management, incentives to not critique Enron or its managers, formulaic auditing standards, and a lack of expertise and resources among the auditor’s oversight committee \cite{id20}.

In light of such potential failure modes, we propose the ASPIRE framework for ensuring effective external scrutiny, illustrated in Figure 1, comprising six requirements: Access, Searching attitude, Proportionality, Independence, Resources, and Expertise. For policy recommendations building on this framework, see Appendix I.

\begin{center}
    \centering
    \includegraphics[width=1\linewidth]{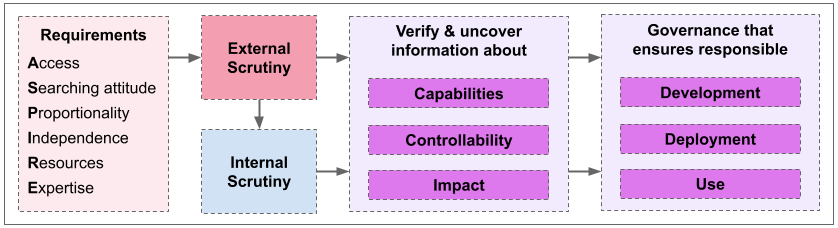}
    {Figure 1. External Scrutiny Requirements and Goals}
    \label{fig:enter-label}
\end{center}

\subsection{Access}
Scrutinizers will require access to AI models and information about their development and use in order to evaluate them. Specifically, they will need access to, among other things: “base models,” “model families,” the components of a deployed AI system, background information on the model, third-party data on the model’s impacts, and the ability to fine-tune the model. For further discussion see Appendix II.

However, providing wider access can increase the risk of unintentional and irreversible model proliferation \cite{os}. This proliferation can limit the tools available to govern AI in a variety of ways, as well as place  dangerous capabilities in the hands of malicious actors. Proliferation can occur through the model itself being leaked or stolen, or through the spread of information about it, such as its training methods, size, or capabilities, making it easier for malicious actors to create a similarly capable model \cite{id19-2}. Policymakers and AI developers must therefore balance providing access with ensuring information security. One such approach is “structured access”, facilitating “controlled, arm’s length interactions with AI systems” through cloud-based interfaces \cite{id37}. 

\subsection{Searching Attitude}
Formulaic approaches to scrutinize models (such as benchmarks) have the benefit of being easily standardized, and while many are highly informative, they are insufficient to comprehensively identify risk \cite{id29, id30}. In addition to formulaic methods, external scrutiny will require a \textit{searching attitude.}

For many risks, like LLMs increasing bioweapons proliferation, benchmarks are still in development. Many benchmarks are rendered obsolete quickly as AI systems improve \cite{id31-1,id31-2}. Further, formulaic methods can be more easily gamed, similar to how Volkswagen designed their cars to meet emissions standards only while being tested by regulators \cite{id32}. They can also fail to accurately simulate real world conditions. As such, they might fail to elicit dangerous behavior or overlook broader societal implications \cite{id1}.

Scrutinizers must therefore be actively incentivized to discover issues. In addition, they need appropriate legal protections against liability or retaliation for good faith scrutiny \cite{Chowdhury}. They must be creative and exploratory in their attempts to “break” models – to elicit some kind of undesirable behavior – and to “stretch” them – to elicit upper-bound capabilities. They must go beyond testing for known risks or capabilities; they must try to discover unknown capabilities \cite{Brundage}. Ultimately, external scrutinizers might be better understood as scientists rather than auditors – trying to push the frontier of understanding forward, rather than evaluating a system against best practice.

\subsection{Proportionality to risk}
The level of scrutiny an LLM faces should scale with the level of risk it poses. Proxies for risk include on one hand those relating to the development process, such as the model's capabilities, its novelty of design, susceptibility to accidents, and potential for misuse and on the other hand, by the LLMs deployment characteristics, including intended domain of application and the number of people likely to be affected. In aviation, for instance, alterations to an existing aircraft design face less scrutiny during certification than entirely new designs do \cite{id44}.

More capable systems are more likely to possess significant dangerous capabilities, with accompanying misuse risks. Further, even when benign, higher performing systems will be used more widely and relied upon for more important decisions, increasing the stakes of accidents. As it is impossible to know the exact capabilities of a system before those capabilities are evaluated, regulators should use predictions of a model’s capabilities based on previous models and be able to increase or decrease scrutiny as the model’s true capabilities become apparent.

More novel systems are less well understood and therefore more likely to produce unknown risks. For instance, a system might be considered more novel if it uses more data and compute than previous systems, or a new training method. Certain safety-critical applications of AI (such as aviation, medicine, or law enforcement) will tend to be higher risk. Similarly, a model which might affect many individuals should also be considered higher risk, all else being equal.

\subsection{Independence}
The quality of external scrutiny is partially determined by the independence of scrutinizers.  As argued by Raji et al. \cite{id7-1}, to avoid poor incentives and guarantee sufficient independence, the AI developer must give up some control over the scrutiny process. Specifically, they must relinquish some control over decisions related to:

\begin{itemize}
    \item \textbf{Selection and compensation. }How are scrutinizers selected to evaluate a particular model, and by whom? How are they compensated? An AI developer who controls the selection and compensation of scrutinizers can apply pressure to them and incentivize friendly treatment.
    \item \textbf{Scope and methods.} What kinds of questions will scrutinizers be answering? What methods will they employ to answer these questions? An AI developer that controls the scope of external scrutiny could mark important questions as off-limits.
    \item \textbf{Access.} What level of access is given to which scrutinizers, and who makes that decision? Since access is a key input for external scrutiny, whoever controls how access is granted holds de-facto veto power over any scrutinizing activity.
    \item \textbf{Post-scrutiny actions.} What is done with the results of scrutiny? Who are the results reported to and how? If an AI developer can bar a scrutinizer from informing relevant actors, then it can block information from being used effectively.
\end{itemize}

\subsection{Resources}
External scrutinizers need to be given the time, financial, and the computational resources necessary to carry out their task effectively. 
\begin{itemize}
    \item \textbf{Time.} Rushing scrutiny can lead to poor risk assessments. The Boeing 737 MAX crashes were caused in part by a rushed certification process attempting to keep up with competitive airplane delivery schedules \cite{id42}. GPT-4, a recent frontier LLM, received six months of pre-deployment evaluations \cite{id43}. As model risks or the complexity of evaluations increase, time for pre-deployment evaluations should increase accordingly. For reference, new pharmaceuticals or aircraft designs can receive several years of testing and evaluation before becoming commercially available \cite{id44, Sertkaya}.  
    \item \textbf{Financial resources and compute.} Scrutinzers should be compensated at rates competitive enough to attract top experts in relevant fields. Further, AI developers should make sure that scrutinzers have access to sufficient computing resources, for example by offering API credits. 
\end{itemize}

\subsection{Expertise}
 External scrutinizers must have sufficient expertise. Given LLMs’ general-purpose nature and complexity \cite{id25}, no single individual or team could possess all the wide-ranging expertise and perspectives needed to answer all the relevant questions \cite{id19-1}. For external scrutiny to be effective, it must represent and incorporate “a diversity of institutions, cultures, demographic groups, languages, and disciplines to be able to critically examine foundation models from different perspectives" \cite{id27}. 
 Further, deep expertise will be required in a number of areas. For instance, to assess biosecurity risks, expertise countering biological weapons proliferation is likely necessary, as well an understanding of what harm different pathogens can cause. This expertise may be possessed by only a few government actors, and is therefore difficult to hire \cite{id5, anthropic}.

\section{External Scrutiny in the AI Lifecycle}

Just as risks from LLMs can emerge at various stages of the AI lifecycle, external scrutiny must be applied throughout these phases. In this section, we illustrate the role external scrutiny can play across the development, pre-deployment and post-deployment stages.

\begin{center}
    \centering
    \includegraphics[width=1\linewidth]{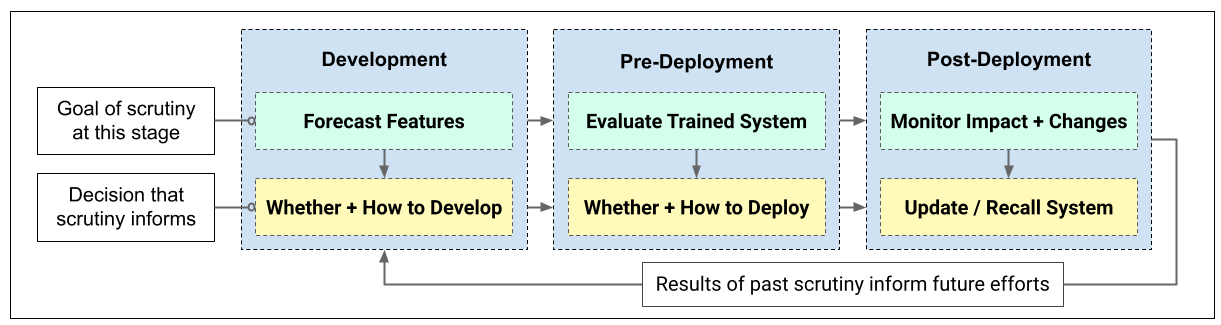}
    {Figure 2. External Scrutiny throughout the AI Lifecycle.}
    \label{"Lifecycle view"}
\end{center}

\subsection{Development}
The results of external scrutiny should inform decisions about frontier LLM development such as whether to train the model in the first place and what information security measures might be warranted. By identifying safety issues early in the lifecycle, they may be less costly to fix \cite{id49-1}. Moreover, high information security standards may need to be adopted for models with significant misuse potential, as they could be stolen after development and before deployment \cite{id52}. 

External scrutinizers should support and assess AI developer’s adherence to their public commitments. One class of such commitments are responsible scaling policies, which outline “conditions under which it would be too dangerous to continue deploying AI systems and/or scaling up AI capabilities until protective measures improve \cite{arc,anr}.” External scrutinizers could predict a model’s capabilities and controllability by studying similar, smaller models, or by inspecting the training data \cite{id19-1}. During training, they should regularly evaluate the model to inform whether to continue or adjust the development process. Thereby, external scrutinizers could assess whether a model is more capable than the AI developer’s safety measures can handle.  

\subsection{Pre-deployment}
The results of external scrutiny should inform decisions on whether and how to deploy the LLM. They can inform these decisions by simulating how malicious actors could make use of the LLM, for instance by assessing the models’ ability to assist in acquiring biological weapons. They can also stress-test the safeguards placed on the LLM and find areas where it could engage in harmful behavior or cause accidents. Further, they can help developers identify relevant information that should be reported to downstream users, helping them determine whether the model is suitable for a particular use-case, and preparing for ways it might cause harm \cite{instance}. 

\subsection{Post-deployment}
Once models are deployed, external scrutinizers can play an important role in assessing changes in system capabilities and controllability, as well as its downstream impacts. Historically, many model capabilities were only discovered after deployment by users, such as through prompt engineering or giving the model access to tools by e.g. combining it with other software \cite{id19-1}. Models which appeared harmless at the time of first deployment might prove more dangerous years later as users discover new capabilities. As such, for models deployed via API, external scrutiny should inform decisions about what updates to safeguards are needed or whether to restrict access to the model, such as by recalling it \cite{obrien}. 

\newpage

\appendix

\newpage
\section*{Appendix I: Policy Recommendations}
\addcontentsline{toc}{section}{Appendix I: Policy Recommendations}
Building an external scrutiny ecosystem that can contribute to public accountability in a meaningful way will be a difficult and lengthy task. However, we believe that there are concrete actions across the ASPIRE framework that policymakers can take into account now when designing new standards, regulations, or legislation.

\subsection*{Access}
\begin{itemize}[left=0em, after=\vspace{-0.6\baselineskip}]
    \item \textbf{Mandate sufficient and secure model access} to qualified external scrutinizers and develop robust norms and criteria to provide such access, such as a scrutinizer accreditation or vetting processes. Qualification might depend on the scrutinizer’s expertise, independence, and trustworthiness to not leak dangerous information.
    \item \textbf{Develop a research API and provide access to third-party data}. This research API could be developed by AI companies, a government body, or through a public-private partnership. It could be managed by a third party, such as the National AI Research Resource in the US.
    \item \textbf{Support the development of structured access} \textbf{tools} and privacy-enhancing technologies.
\end{itemize}

\subsection*{Searching attitude}
\begin{itemize}[left=0em, after=\vspace{-0.6\baselineskip}]
    \item \textbf{Ensure scrutinizers are not subject to undue liability or retaliation}, e.g., by creating safe harbors. Otherwise, fear of repercussions for honest mistakes might discourage exploratory scrutiny and bias the scrutinizer towards check-list, compliance-focused approaches to avoid liability \cite{Chowdhury}.
    \item \textbf{Promote competition among scrutinizers} for instance by engaging multiple independent teams to elicit dangerous behavior of a model, in an "adversarial audit". Ideally, compensation should be designed to reward teams that successfully expose flaws.
    \item \textbf{Have standards focus on identifying key objectives for evaluation,} rather than prescribing methods. For example, defining specific undesirable and desirable features of AI models, including worrisome dangerous capabilities, can help steer scrutinizer efforts. This can allow for a more flexible, future-proofed approach that adapts to evolving scrutiny practices while maintaining a targeted assessment.

\end{itemize}

\subsection*{Proportionality}
\begin{itemize}[left=0em, after=\vspace{-0.6\baselineskip}]
    \item \textbf{Scale the intensity of mandated external scrutiny in proportion to the risks} the model poses. This could be guided by a classification scheme, analogous to biosafety levels, which necessitate varying degrees of scrutiny and regulatory oversight based on the level of risk they pose \cite{anr}. 
    
\end{itemize}

\subsection*{Independence}
\begin{itemize}[left=0em, after=\vspace{-0.6\baselineskip}]
    \item \textbf{Develop standards of independence }for external scrutinizers that take into account how decisions on selection and compensation, scope and methods, access, and post-scrutiny actions are made.
    \item \textbf{Reduce the administrative burden for external scrutiny}, e.g., by promoting the development of standardized contracts to decrease red tape and ensuring companies are not able to impose unreasonable NDA requirements. 
    \item \textbf{Establish oversight mechanisms} that can hold scrutinizers accountable, e.g., by setting and monitoring standards on scrutiny, such as rules about conflict of interest.
\end{itemize}

\subsection*{Resources}
\begin{itemize}[left=0em, after=\vspace{-0.6\baselineskip}]
    \item \textbf{Ensure scrutiny is sufficiently resourced}, i.e., AI developers should provide enough financial and computational resources for scrutinizers, and not rush their development or deployment timelines. Current frontier LLMs should see at least six months of external scrutiny or red-teaming before deployment. As potential risks increase, scrutiny time should increase as well. 
\end{itemize}

\subsection*{Expertise}
\begin{itemize}[left=0em]
    \item \textbf{Fund research programs }aimed at increasing capacity to evaluate LLMs outside AI developers. This could include training of diverse research talent, e.g., through the National AI Research Resource.
    \item \textbf{Provide government expertise} for scrutiny in areas where it is particularly necessary, such as in assessments relevant to national security.
\end{itemize}

\newpage

\section*{Appendix II: Access Requirements}

There are many different kinds of access that scrutinizers will require to evaluate frontier LLMs. For more details, see Bucknall et al \cite{ben}. Seeing as such access would risk revealing proprietary or sensitive information, it should be designed such as to minimize undesirable information or model leakage. Scrutinizers will require sufficient access to:

\begin{itemize}
    \item \textbf{Efficient sampling.} Scrutinizers must be able to interact with the model, querying it and seeing its outputs. Scrutinizers will likely need to sample the AI model many times, so the ability to do so in an automated and systematic manner is important. Scrutinizers should have more functionality than that, as model behavior can be improved and modified by implementing different sampling algorithms. Further, scrutinizers should have access to the logits and derived probabilities of a model’s output as they are crucial for calculating cross-entropy loss and perplexity – two standard measures of a model’s performance.
    \item \textbf{Fine-tune the AI model.} Fine-tuning is a process that alters an AI model to exhibit new capabilities and tendencies. Fine-tuning is necessary to allow red-teamers to understand the full range of capabilities and behavior the model possesses. 
    \item \textbf{The ``base model”.} Access to versions of the model that lack safety mitigations, like fine-tuning, is useful to understand latent capabilities within the model, and to understand what the model would be capable of if users are ever able to disable its safeguards. However, base model access can be facilitated through an API and does not require giving researchers access to “model weights”. 
    \item \textbf{Model families and previous versions.} Models often come in families that vary along one or more dimensions such as amount of data, training compute, and the type and extent of fine-tuning. Access to such model families is especially useful for studying scaling laws, which important for predicting the capabilities of future models. In addition, researchers benefit from studying older models after newer models are released, though these older models are often made unavailable. 
    \item \textbf{Model internals. } To understand why models behave a certain way and make better predictions about their behavior, scrutinizers may need access to model internals, including e.g. model activations, attention, and embeddings. Though it should be noted that current internals-based approaches to model evaluation are in their infancy. 
    \item \textbf{Training data}. The data used to train a model may have significant influence over its eventual behavior and failure modes. By studying the training data of a model, scrutinizers may be better able to predict if a model will exhibit dangerous capabilities. Access to the training data itself may not be necessary if scrutinizers have a way to understand the data sources, data cleaning decisions, and access to metadata about the data composition. 
    \item \textbf{All the components of the deployed AI system.} Deployed AI systems typically combine a core model with smaller models and other software components and tools. Where such components exist, scrutinizers benefit from accessing them to understand how the AI system is likely to behave in production.
    \item \textbf{Third-party data on the system’s impact.} To understand an LLM’s impact, it will be necessary to access data from sources other than just the AI developer. For instance, to understand the impact of AI generated disinformation on social media, some kind of privacy-preserving data sharing agreement could be established with social media companies \cite{id36}.  
    \item \textbf{Model information.} Researchers will find it useful to know various kinds of meta-information about the model, including the input data, underlying training algorithm, internal testing results and amount of training compute. Developers publish such information in model or system cards \cite{id43,id17-1,id17-2,instance}, though there are privacy and proliferation concerns associated with making such information publicly available.
\end{itemize}

\newpage
\section*{Appendix III: Social Impacts Statement}

\textbf{Aim.} Our paper aims to inform ongoing policy discussions on how model evaluations, auditing, red-teaming, and independent researcher access can contribute to public accountability. We hope to provide decision-makers with a framework to assess key design considerations for external scrutiny to ensure that AI policy is well-informed, capable of serving the public interest and of holding extremely powerful AI companies accountable. We particularly hope that by raising the salience of external scrutiny as a policy tool, and by articulating how it can be done effectively, audits, model evaluations, and red teaming can pave the way for broader stakeholder involvement in governing frontier LLMs. 

\textbf{Uncertainties. }While we believe there is value to this framework as a high-level design tool, more granular, concrete decisions will need to be made to implement external scrutiny, and those granular decisions are beyond the scope of this paper. Although we would like to provide more concrete recommendations for policymakers, they crucially depend on the details of a situation and political context. Therefore, we have erred on the side of advocating for what seems clearly beneficial across many situations at the expense of being less concrete. 

\textbf{Limitations.}  It is important to acknowledge that even well-designed external scrutiny has its limitations. For instance, external scrutiny is unlikely to identify all the important risks in AI, and too much faith might be placed in the results of scrutiny, leading to a sense of false security. Further, external scrutiny will only have an impact insofar as it informs and changes important decisions, which may require e.g. regulators and the public to have more levers of influence over AI developers. As such, external scrutiny must be seen as only one tool in the AI governance toolbox. Further, we would like to underscore that our focus on the role external actors should play does not exempt AI developers from being ultimately responsible for guaranteeing their systems are safe. 

\newpage

\bibliographystyle{IEEEtran}
\bibliography{main}

%%%%%%%%%%%%%%%%%%%%%%%%%%%%%%%%%%%%%%%%%%%%%%%%%%%%%%%%%%%%

\end{document}